\documentclass[aps,prd,showpacs,nofootinbib,tightenlines]{revtex4}
\usepackage{amsmath}
\usepackage{amssymb}
\usepackage{epsfig}
\usepackage{graphicx}
\begin{document}

\newcommand{\beq}{\begin{eqnarray}}
\newcommand{\eeq}{\end{eqnarray}}

\newcommand{\non}{\nonumber\\ }
\newcommand{\rmt}{ {\rm T}}
\newcommand{\psl}{ p \hspace{-2.0truemm}/ }
\newcommand{\qsl}{ q \hspace{-2.0truemm}/ }
\newcommand{\epsl}{ \epsilon \hspace{-2.0truemm}/ }
\newcommand{\nsl}{ n \hspace{-2.2truemm}/ }
\newcommand{\vsl}{ v \hspace{-2.2truemm}/ }

\def \ctp{ {Commun. Theor. Phys. }}
\def \epjc{ {Eur. Phys. J. C }}
\def \jhep{ {J. High Energy Phys. }}
\def \jpg{ {J. Phys. G }}
\def \npb{ {Nucl. Phys. {\bf B}}}
\def \plb{ {Phys. Lett. B }}
\def \prd{ {Phys. Rev. D }}
\def \prl{ {Phys. Rev. Lett. }}
\def \ptp{ {Prog. Theor. Phys. }}
\def \zpc{ {Z. Phys. C }}
\def \cpc{ {Chin. Phys. C }}


\title{Direct {\sl CP} asymmetries of three-body $B$ decays in perturbative QCD}
\author{Wen-Fei Wang$^1$}\email{wangwf@ihep.ac.cn}
\author{Hao-Chung Hu$^{2,3}$}\email{hchu@phys.sinica.edu.tw}
\author{Hsiang-nan Li$^{3,4,5}$}\email{hnli@phys.sinica.edu.tw}
\author{Cai-Dian L\"u$^1$}\email{lucd@ihep.ac.cn}
\affiliation{$^{1}$Institute of High Energy Physics and Theoretical Physics
Center for Science  Facilities, Chinese Academy of Sciences, Beijing
100049, People's Republic of China,}
\affiliation{$^{2}$Department of Physics, National Taiwan
University, Taipei, Taiwan 106, Republic of China,}
\affiliation{$^{3}$Institute of Physics, Academia Sinica, Taipei,
Taiwan 115, Republic of China,}
\affiliation{$^{4}$Department of Physics, National Tsing-Hua
University, Hsinchu, Taiwan 300, Republic of China,}
\affiliation{$^{5}$Department of Physics, National Cheng-Kung
University, Tainan, Taiwan 701, Republic of China}

\date{\today}

\begin{abstract}
We propose a theoretical framework for analyzing three-body hadronic
$B$ meson decays based on the perturbative QCD approach. The crucial
nonperturbative input is a two-hadron distribution amplitude
for final states, whose time-like form factor and
rescattering phase are fit to relevant experimental data. Together
with the short-distance strong phase from the $b$-quark decay
kernel, we are able to make predictions for direct {\sl CP} asymmetries in,
for example, the $B^\pm \to \pi^+ \pi^- \pi^\pm$ and
$\pi^+\pi^-K^\pm$ modes, which are consistent with the LHCb data in
various localized regions of phase space.
Applications of our formalism to other three-body
hadronic and radiative $B$ meson decays are mentioned.

\end{abstract}


\maketitle

Three-body hadronic $B$ meson decays have been studied for many years
\cite{CL02,CY02,FPP04,BIL13}. They attracted much attention recently, after
the LHCb Collaboration measured sizable direct {\sl CP} asymmetries in localized regions
of phase space \cite{LHCb1,LHCb2,LHCb3}, such as
\begin{eqnarray}
A_{CP}^{\rm reg}(\pi^+\pi^-\pi^+) = 0.584 \pm 0.082 \pm 0.027 \pm 0.007,
\end{eqnarray}
for $m^2_{\pi^+\pi^- \rm high} > 15$ GeV$^2$ and $m^2_{\pi^+\pi^-\rm low} < 0.4$ GeV$^2$, and
\begin{eqnarray}
A_{CP}^{\rm reg}(\pi^+\pi^-K^+) = 0.678 \pm 0.078 \pm 0.032 \pm 0.007, \label{kpp}
\end{eqnarray}
for $m^2_{K^+\pi^- \rm high} < 15$ GeV$^2$ and $0.08 < m^2_{\pi^+\pi^-\rm low} < 0.66$ GeV$^2$.
Theoretical attempts to understand these data were made: The above
{\sl CP} asymmetries were attributed to the interference between a light scalar and
intermediate resonances in \cite{ZGY13}; the relations among the above CP asymmetries in the U-spin
symmetry limit were examined in \cite{BGR13}; SU(3) and  U-spin symmetry breaking effects
were included in the amplitude parametrization in \cite{XLH13}; in \cite{CT13} the
non-resonant contributions were parameterized in the framework of heavy meson chiral
perturbation theory \cite{LLW92}; and the resonant contributions were
estimated by means of the usual Breit-Wigner formalism.

Viewing the experimental progress, it is important to construct a corresponding framework
based on the factorization theorem, in which perturbative evaluation can be performed
systematically with controllable nonperturbative inputs. Motivated by its theoretical
self-consistency and phenomenological success, we shall generalize the perturbative
QCD (PQCD) approach \cite{KLS,Lu:pqcd} to three-body hadronic $B$ meson decays.
A direct evaluation of hard $b$-quark decay kernels, which contain two virtual gluons
at leading order (LO), is not practical because of the enormous number of diagrams. Besides, the
contribution from two hard gluons is power-suppressed and is not important. In this region
all three final-state mesons carry momenta of $O(m_B)$, and all three pairs of them have
invariant masses of $O(m_B^2)$, $m_B$ being the $B$ meson mass. The dominant
contribution comes from the region, where at least one pair of light mesons has an
invariant mass below $O(\bar\Lambda m_B)$ \cite{CL02}, $\bar\Lambda=m_B-m_b$
being the $B$ meson and $b$ quark mass difference. The configuration
involves two energetic mesons almost collimating to each other, in which
the dynamics associated with the pair of mesons
can be factorized into a two-meson distribution
amplitude $\phi_{h_1h_2}$ \cite{MP}. It is evident that $\phi_{h_1h_2}$ appropriately 
describes the nonperturbative dynamics of a two-meson system in the 
localized region of phase space, say, $m^2_{\pi^+\pi^-\rm low} < 0.4$ GeV$^2$.

\begin{figure}[tbp]
\centerline{\epsfxsize=13cm \epsffile{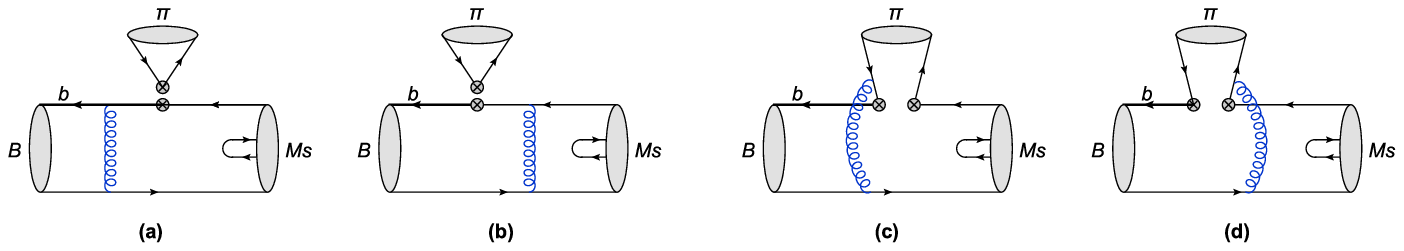}}
\caption{Single-pion emission diagrams for the
$B^+\to \pi^+\pi^-\pi^+$ decay, where $Ms$ stands for the pion pair.}\label{fig-fig1}
\vspace{0.5cm}
\centerline{\epsfxsize=13cm \epsffile{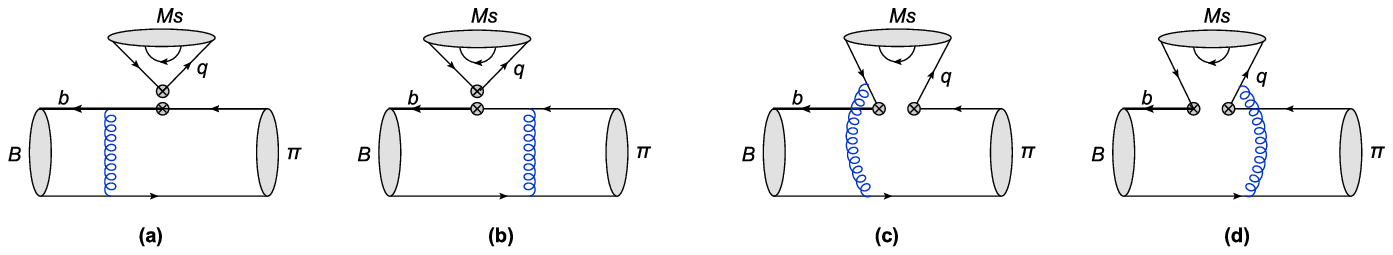}}
\caption{Two-pion emission diagrams, where $q$ denotes a $u$ or $d$ quark.}\label{fig-fig2}
\vspace{0.5cm}
\centerline{\epsfxsize=13cm \epsffile{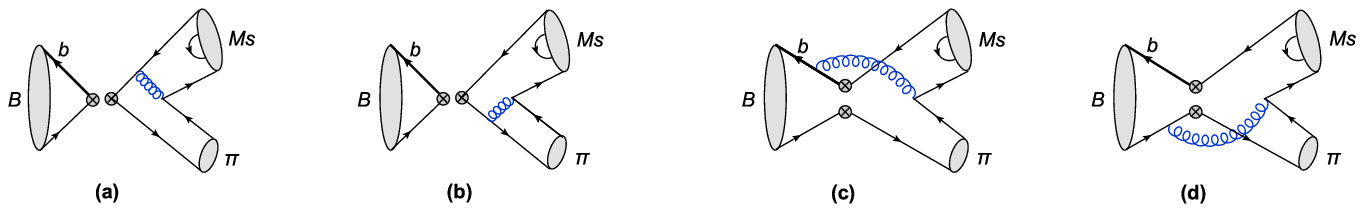}}
\caption{Annihilation diagrams.}\label{fig-fig3}
\vspace{0.5cm}
\centerline{\epsfxsize=13cm \epsffile{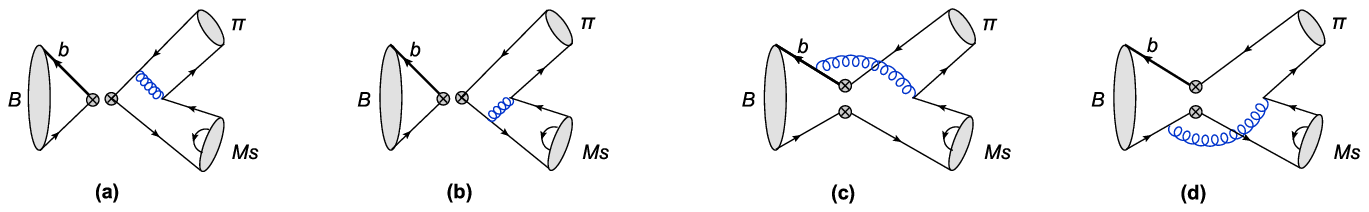}}
\caption{More annihilation diagrams.}
\label{fig-figs}
\end{figure}

With the introduction of a two-meson distribution amplitude, the
LO diagrams for three-body hadronic $B$ meson decays reduce to
those for two-body decays, as displayed in Figs.~\ref{fig-fig1}-\ref{fig-figs}.
The PQCD factorization formula for a $B\to h_1h_2h_3$ decay amplitude is then
written as \cite{CL02}
\begin{eqnarray}
\mathcal{A}=\phi_B\otimes H\otimes \phi_{h_1h_2}\otimes\phi_{h_3},
\end{eqnarray}
where the hard kernel $H$ contains only a single hard gluon.
The $B$ meson ($h_1$-$h_2$ pair, $h_3$ meson) distribution amplitude
$\phi_B$ ($\phi_{h_1h_2}$, $\phi_{h_3}$) absorbs nonperturbative
dynamics characterized by the soft scale $\bar\Lambda$
(the invariant mass of the meson pair, the $h_3$ meson mass).
Figure~\ref{fig-fig1} involves the transition of the $B$ meson into
two light mesons. The amplitude from Fig.~\ref{fig-fig2} is expressed as a
product of a heavy-to-light form factor and a time-like light-light form
factor in the heavy-quark limit. In Figs.~\ref{fig-fig3} and \ref{fig-figs},
a $B$ meson annihilates completely, and three light mesons are produced.

Take Fig.~1(a) for the $B^+\to \pi^+\pi^-\pi^+$ decay as an example,
in which the $B^+$ meson momentum $p_B$, the total momentum  $p=p_1+p_2$
of the pion pair, and the momentum $p_3$ of the second $\pi^+$ meson are chosen,
in light-cone coordinates, as
\beq\label{mom-pBpp3}
& &p_B=\frac{m_B}{\sqrt2}(1,1,0_\rmt),~\quad p=\frac{m_B}{\sqrt2}(1,\eta,0_\rmt),~\quad
p_3=\frac{m_B}{\sqrt2}(0,1-\eta,0_\rmt),
\eeq
with the variable $\eta=\omega^2/m^2_B$,
$\omega^2=p^2$ being the invariant mass squared.
The momenta $p_1$ and $p_2$ of the $\pi^+$ and $\pi^-$ mesons
in the pair, respectively, have the components
\beq
p^+_1=\zeta\frac{m_B}{\sqrt2}, \quad p^-_1=(1-\zeta)\eta\frac{m_B}{\sqrt2},
\quad p^+_2=(1-\zeta)\frac{m_B}{\sqrt2}, \quad p^-_2=\zeta\eta\frac{m_B}{\sqrt2},
\eeq
with the $\pi^+$ meson momentum fraction $\zeta$. The momenta of the spectators
in the $B$ meson, the pion pair, and the $\pi^+$ meson read, respectively, as
\beq\label{mom-B-k}
k_B=\left(0,\frac{m_B}{\sqrt2}x_B,k_{B\rmt}\right), \quad
k=\left(\frac{m_B}{\sqrt2}z,0,k_\rmt\right),\quad
k_3=\left(0,\frac{m_B}{\sqrt2}(1-\eta)x_3,k_{3\rmt}\right).
\eeq

The definitions of the two-pion distribution amplitudes in terms of
hadronic matrix elements of nonlocal quark operators up to twist 3
can be found in \cite{CL02,MP,DGP00}. We parameterize them at the
leading partial waves as
\beq\label{exp-wf-PiPi}
&&\phi^{v,t}_{\pi\pi}(z,\zeta,\omega^2)=\frac{3F_{\pi,t}(\omega^2)}{\sqrt{2N_c}}
z(1-z)(2\zeta-1),\\
&&\phi^{s}_{\pi\pi}(z,\zeta,\omega^2)=\frac{3F_{s}(\omega^2)}{\sqrt{2N_c}}
z(1-z),
\eeq
with the number of colors $N_c$, where the factor $2\zeta-1$ arises from
the Legendre polynomial $P_l(2\zeta-1)$ for $l=1$. The PQCD power counting
indicates the scaling of the vector-current form factor in the asymptotic region,
$F_\pi(w^2)\sim 1/w^2$, and the relative importance of the scalar-current and
tensor-current form factors, $F_{s,t}(w^2)/F_\pi(w^2)\sim m_0^\pi/w$,
where $m_0^\pi=m_\pi^2/(m_u+m_d)$ is the chiral scale associated with the pion,
$m_\pi$, $m_u$, and $m_d$ being the masses of the pion, the $u$ quark, and
the $d$ quark, respectively. To evaluate the nonresonant contribution in the
arbitrary range of $w^2$, we propose the parametrization for the complex time-like
form factors
\begin{eqnarray}
F_\pi(w^2)=\frac{m^2\exp[i\delta^1_1(w)]}{w^2+m^2},\;\;\;\;
F_{t}(w^2)=\frac{m_0^\pi m^2\exp[i\delta^1_{1}(w)]}{w^3+m_0^\pi m^2},\;\;\;\;
F_{s}(w^2)=\frac{m_0^\pi m^2\exp[i\delta^0_{0}(w)]}{w^3+m_0^\pi m^2},
\label{non}
\end{eqnarray}
in which the parameter $m=1$ GeV is determined by the fit to
the experimental data $m_{J/\psi}^2|F_\pi(m_{J/\psi}^2)|^2\sim 0.9$
GeV$^2$ \cite{PDG}, $m_{J/\psi}$ being the $J/\psi$ meson mass. The resultant
$w^2$ dependence of $F_\pi(w^2)$ also agrees with the low-energy data of the
time-like pion electromagnetic form factor for $w<1$ GeV \cite{Whalley2003aaa},
and with the next-to-leading-order (NLO) PQCD calculation \cite{HL13}. The strong
phases $\delta^I_l$ are chosen as the phase shifts for
the $S$ wave ($I=0$, $l=0$) and $P$ wave ($I=1$, $l=1$) of elastic $\pi\pi$
scattering \cite{DGP00} according to Watson's theorem.
We simply parameterize the data of these
strong phases \cite{Proto1973,EM74,KS90} for $2m_\pi<w< 0.7$ GeV as
\begin{eqnarray}
& &\delta^0_0(w)=\pi (w-2m_\pi),\;\;\;\;
\delta^1_1(w)=1.4\pi (w-2m_\pi)^2,
\end{eqnarray}
in which $2m_\pi$ represents the $\pi\pi$ threshold. The increase of $\delta_1^1$ with
$w$ in the above expression is consistent with the NLO PQCD result of
the time-like pion electromagnetic form factor \cite{HL13}.

The $B$ meson, pion, and
kaon distribution amplitudes are the same as those widely adopted in the PQCD
approach to two-body hadronic $B$ meson decays.
We have the $B$ meson distribution amplitude
\beq
\phi_{B}(x,b)&=& N_Bx^2(1-x)^2
\exp\left[-\frac{1}{2}\left(\frac{xm_B}{\omega_B}\right)^2
-\frac{\omega_B^2 b^2}{2}\right],
\eeq
with the shape parameter $\omega_B=0.45\pm0.05$ GeV, and the normalization
constant $N_B=73.67$ GeV being related to the $B$ meson decay constant $f_B=0.21$ GeV via
$\lim_{b\to 0}\int dx  \phi_{B}(x,b)=f_B/(2\sqrt{2N_c})$.
The pion and kaon distribution amplitudes up to twist 3, $\phi_i^A(x)$ and
$\phi_i^{P,T}(x)$ for $i=\pi,K$, are chosen as \cite{refs-pball}
\beq
\phi_i^A(x) &=& \frac{3 f_i}{\sqrt{6} }\, x(1-x) \left [ 1 + a_1
C_1^{3/2}(t) + a_2 C_2^{3/2}(t)+a_4 C_4^{3/2}(t)\right] ,
\label{eq:phipik-a}\\
\phi^P_i(x) &=& \frac{f_i}{2\sqrt{6}}\, \left [ 1 +\left(30\eta_3
-\frac{5}{2}\rho_i^2\right) C_2^{1/2}(t)  -\, 3\left\{
\eta_3\omega_3 + \frac{9}{20}\rho_i^2(1+6a_2 ) \right\} C_4^{1/2}(t)
\right ], \label{eq:phipik-p}
\\
\phi^\sigma_i(x) &=& \frac{f_i}{2\sqrt{6}}\, x(1-x) \left [ 1
+ \left(5\eta_3 -\frac{1}{2}\eta_3\omega_3 -
\frac{7}{20} \rho_i^2 - \frac{3}{5}\rho_i^2 a_2 \right)C_2^{3/2}(t) \right ],
\label{eq:phipik-t}
\eeq
with the pion (kaon) decay constant $f_\pi=0.13$ ($f_K=0.16$) GeV,
the variable $t=2x-1$, the Gegenbauer polynomials
\beq
C_1^{3/2}(t)\, &=&  3\, t \;, \quad
C_2^{1/2}(t)= \frac{1}{2} \left(3\, t^2-1\right), \quad
C_2^{3/2}(t)\, =\, \frac{3}{2} \left(5\, t^2-1\right), \non
C_4^{1/2}(t)\, &=& \, \frac{1}{8} \left(3-30\, t^2+35\, t^4\right), \quad
C_4^{3/2}(t) \,=\, \frac{15}{8} \left(1-14\, t^2+21\, t^4\right),
\label{eq:cii}
\eeq
and the mass ratio $\rho_{\pi(K)}=m_{\pi(K)}/m_0^{\pi(K)}$, where
$m_0^K=m_K^2/(m_s+m_d)$ is the chiral scale associated with the kaon,
$m_K$ and $m_s$ being the masses of the kaon and the $s$ quark, respectively.
The Gegenbauer moments $a^{\pi,K}$ are set to \cite{refs-pball}
\beq
a_1^\pi&=& 0, \quad a_1^K=0.06\pm 0.03, \quad a_2^{\pi,K}=0.25 \pm 0.15, \non
a_4^\pi &=& -0.015, \quad \eta_3^{\pi, K}=0.015, \quad \omega_3^{\pi,K} =-3.
\label{eq:gms}
\eeq
The above set of meson distribution amplitudes corresponds to the
$B\to\pi$ transition form factors at maximal recoil $F_+^{B\pi}(0)=F_0^{B\pi}(0)=0.23$
in LO PQCD, which are consistent with the results derived from other
approaches \cite{refs-pball,bpi}.

The $B^+ \to \pi^+ \pi^- \pi^+$ decay width in the localized region of
$m^2_{\pi^+\pi^-\min}<m^2_{\min} = 0.4$ GeV$^2$ and
$m^2_{\pi^+\pi^-\max} > m^2_{\max}=15$ GeV$^2$ is written as
\begin{eqnarray}
\Gamma=\frac{G_F^2m_B}{512\pi^4}\int_{\eta_{\min}}^{\eta_{\max}} d\eta (1-\eta)
\int_0^{\zeta_{\max}} d\zeta|\mathcal{A}|^2,
\end{eqnarray}
with the Fermi constant $G_F=1.16639 \time 10^{-5}$ GeV$^{-2}$ and the bounds
\begin{eqnarray}
\eta_{\max}=\frac{m^2_{\min}}{m_B^2},\;\;\;\;
\eta_{\min}=\frac{4m^2_\pi}{m_B^2},\;\;\;\;
\zeta_{\max}=1-\frac{m^2_{\max}}{(1-\eta)m_B^2},
\end{eqnarray}
where the upper bound $\zeta_{\max}$ is derived from the invariant mass
squared $(p_2+p_3)^2$. The contributions from all the diagrams in
Figs.~\ref{fig-fig1}-\ref{fig-figs} to the decay amplitude $\mathcal{A}$ are
collected in the Appendix. The corresponding formulas for the $B^+ \to \pi^+ \pi^- K^+$
decay can be obtained straightforwardly.

Employing the input parameters $\Lambda^{(f=4)}_{\overline{MS}}=0.25$ GeV,
$m_{\pi^\pm}=0.1396$ GeV, $m_{K^\pm}=0.4937$ GeV, $m_{B^\pm}=5.279$ GeV
\cite{PDG,prd76-074018}, and the Wolfenstein parameters in \cite{PDG}, we derive
the direct {\sl CP} asymmetries in the region of
$m^2_{\pi^+\pi^- {\rm low}}<0.4$ GeV$^2$ and
$m^2_{\pi^+\pi^-\;{\rm or}\;K^+\pi^- {\rm high}}>15$ GeV$^2$,
\begin{eqnarray}
A_{CP}(B^\pm\to\pi^+\pi^-\pi^\pm)&=&0.519^{+0.124}_{-0.219}(\omega_B)^{+0.108}_{-0.091}(a_2^\pi)
^{+0.027}_{-0.032}(m^\pi_0), \\
A_{CP}(B^\pm\to \pi^+\pi^-K^\pm)&=&-0.018^{+0.024}_{-0.044}(\omega_B)
^{+0.006}_{-0.009}(a_2^\pi\;\&\;a_2^K)^{+0.002}_{-0.003}(m_0^\pi\;\&\;m_0^K).\label{kpp2}
\end{eqnarray}
The first and second errors come from the variation of $\omega_B=0.45\pm0.05$ GeV
and $a_2^{\pi,K}=0.25\pm0.15$, respectively, and the third errors are induced
by $m_0^\pi=1.4\pm0.1$ GeV and $m_0^K=1.6\pm0.1$ GeV.
The uncertainties caused by the variation of the Wolfenstein parameters $\lambda, A, \rho,
\eta$, and of the Gegenbauer moment $a^K_1=0.06\pm0.03$ are very small,
and have been neglected. While the decay widths are quadratically proportional to
the decay constants $f_B, f_\pi$ and/or $f_K$, the {\sl CP} asymmetries
are independent of them.

Obviously, our prediction for $A_{CP}(B^\pm\to\pi^+\pi^-\pi^\pm)$ agrees
well with the LHCb data. Since the emission contribution and the imaginary
annihilation contribution depend on the $B$ meson distribution amplitude
in different ways, the variation of $\omega_B$ explores the relevance of
the short-distance strong phase from the $b$-quark decay kernel. The
sensitivity of the predicted {\sl CP} asymmetries to $\omega_B$ then implies the
importance of this strong phase. As the $P$-wave rescattering phase
associated with the pion electromagnetic form factor decreases to half, the
predicted CP asymmetries are also reduced by half. The change of the phases
associated with the scalar and tensor form factors does not modify the {\sl CP}
asymmetries much. Therefore, we conclude that the short-distance and
long-distance $P$-wave strong phases are equally crucial for the direct {\sl CP} asymmetries
in the localized region of phase space. The LHCb data in Eq.~(\ref{kpp}) are
dominated by the resonant channel $B^\pm\to \rho^0 K^\pm$. It is encouraging
that the data confirm the NLO PQCD prediction
$A_{CP}(B^\pm\to \rho^0 K^\pm)=0.71^{+0.25}_{-0.35}$ \cite{LM06}. We have
checked that our prediction in Eq.~(\ref{kpp2}) for the localized region of phase space
is consistent with the LHCb data in Fig.~2 of \cite{LHCb1}.
Moreover, we have predicted larger
$A_{CP}(B^\pm\to\pi^+\pi^-\pi^\pm)=0.631$ in the region of
$m^2_{\pi^+\pi^- {\rm low}}<0.4$ GeV$^2$ and
$m^2_{\pi^+\pi^- {\rm high}}>20.5$ GeV$^2$ for the central values of
the input parameters, which also matches the data \cite{LHCb2}.

In this paper we have proposed a promising formalism for three-body
hadronic $B$ meson decays based on the PQCD approach. The calculation
is greatly simplified with the introduction of the nonperturbative two-hadron
distribution amplitude for final states.
The time-like form factors and the rescattering phases involved
in the two-pion distribution amplitudes have been fixed by experiments, and
the $B$ meson, pion, and kaon distribution amplitudes are the same as
in the previous PQCD analysis of two-body hadronic $B$ meson decays.
Without any free parameters, our results for $A_{CP}(B^\pm\to\pi^+\pi^-\pi^\pm)$
and $A_{CP}(B^\pm\to \pi^+\pi^-K^\pm)$ accommodate well
the recent LHCb data in various localized regions of phase space. It has been
observed that the short-distance strong phase from the $b$-quark decay
kernel and the final-state rescattering phase are equally important for
explaining the measured direct {\sl CP} asymmetries. The success
indicates that our formalism has potential applications to other three-body
hadronic and radiative $B$ meson decays \cite{CL04}, if phase shifts from
meson-meson scattering can be derived in nonperturbative methods
\cite{LZL,Doring:2013wka}.

\begin{acknowledgments}
We thank Wei Wang  for helpful discussions. This work was partly supported by
the National Science Council of R.O.C. under Grant No. NSC-101-2112-M-001-006-MY3,
by the National Center for Theoretical Sciences of R.O.C., and by the National
Science Foundation of China under Grants No. 11375208, No. 11228512 and No. 11235005.

\end{acknowledgments}

\appendix
\section{Decay amplitudes}\label{sec-da}

In this appendix we present the PQCD factorization formulas for
the diagrams in Figs.~\ref{fig-fig1}-\ref{fig-figs}.
The sum of the contributions from Figs.~\ref{fig-fig1}(a) and \ref{fig-fig1}(b) gives
\beq\label{DA-fig1-ab}
\mathcal{A}_{1(a,b)}&=&V^*_{ub}V_{ud}
F^{LL}_{B\to\pi\pi} -V^*_{tb}V_{td}\left( F^{\prime LL}_{B\to\pi\pi}
+ F^{SP}_{B\to\pi\pi} \right),
\eeq
where the amplitudes for the $B$ meson transition into two pions are written as
\beq\label{exp-FLLPP}
F^{LL}_{B\to\pi\pi}&=&8\pi C_F m^4_B f_\pi\int dx_B dz\int b_B db_B b db \phi_B(x_B,b_B)(1-\eta)\non
&&\times\bigg\{\left[\sqrt{\eta}(1-2z)(\phi_s+\phi_t)+(1+z)\phi_v \right]
a_1(t_{1a})E_{1ab}(t_{1a})h_{1a}(x_B,z,b_B,b)\non
&&+\sqrt{\eta}\left(2\phi_s-\sqrt\eta\phi_v \right)a_1(t_{1b})E_{1ab}(t_{1b})
h_{1b}(x_B,z,b_B,b)
\bigg\},\\
F^{\prime LL}_{B\to\pi\pi}&=&F^{LL}_{B\to\pi\pi}|_{a_1\to a_3}\\
\label{exp-FSPPP}
F^{SP}_{B\to\pi\pi}&=&-16\pi C_F m^4_B r f_\pi \int dx_B dz\int b_B db_B b db \phi_B(x_B,b_B)\non
& &\times\bigg\{\left[\sqrt{\eta}(2+z)\phi_s-\sqrt{\eta}z\phi_t+(1+\eta(1-2z))\phi_v \right]
a_5(t_{1a})E_{1ab}(t_{1a})h_{1a}(x_B,z,b_B,b)\non
&&+\left[2\sqrt{\eta}(1-x_B+\eta)\phi_s+(x_B-2\eta)\phi_v \right]a_5(t_{1b})E_{1ab}(t_{1b})
h_{1b}(x_B,z,b_B,b)
\bigg\},
\eeq
with $r=m^\pi_0/m_B$ and $\phi_{s,t,v}\equiv\phi_{s,t,v}(z,\zeta,\omega^2)$.
The Wilson coefficients in the above expressions are defined as
$a_1=C_1/N_c+C_2$, $a_3=C_3/N_c+C_4+C_9/N_c+C_{10}$, and
$a_5=C_5/N_c+C_6+C_7/N_c+C_8$.
The spectator diagrams in Figs.~\ref{fig-fig1}(c) and \ref{fig-fig1}(d) lead to
\beq\label{DA-fig1-cd}
\mathcal{A}_{1(c,d)}=V^*_{ub}V_{ud}M^{LL}_{B\to\pi\pi}
-V^*_{tb}V_{td}\left(
M^{\prime LL}_{B\to\pi\pi}+M^{LR}_{B\to\pi\pi} \right),
\eeq
with the amplitudes
\beq\label{exp-MLLPP}
M^{LL}_{B\to\pi\pi}&=&32\pi C_F m^4_B/\sqrt{2N_c} \int dx_B dz dx_3\int b_B db_B b_3 db_3
\phi_B(x_B,b_B)\phi^A_\pi(1-\eta)\non
&&\times\bigg\{\left[\sqrt{\eta}z(\phi_s+\phi_t)+((1-\eta)(1-x_3)-x_B+z\eta)\phi_v \right]
C_1(t_{1c}) E_{1cd}(t_{1c})h_{1c}(x_B,z,x_3,b_B,b_3)\non
&&-\left[z(\sqrt{\eta}(\phi_s-\phi_t)+\phi_v )+(x_3(1-\eta)-x_B)\phi_v \right]C_1(t_{1d}) E_{1cd}(t_{1d})
h_{1d}(x_B,z,x_3,b_B,b_3)
\bigg\},\\
M^{\prime LL}_{B\to\pi\pi}&=&M^{LL}_{B\to\pi\pi}|_{C_1\to a_9}\\
\label{exp-MLRPP}
M^{LR}_{B\to\pi\pi}&=&32\pi C_F r m^4_B/\sqrt{2N_c}\int dx_B dz dx_3\int b_B db_B b_3 db_3
\phi_B(x_B,b_B)\non
&&\times\bigg\{\bigg[\sqrt{\eta}z(\phi^P_\pi+\phi^T_\pi)(\phi_s-\phi_t)
+\sqrt{\eta}((1-x_3)(1-\eta)-x_B)(\phi^P_\pi-\phi^T_\pi)\non
&&\times(\phi_s+\phi_t)-((1-x_3)(1-\eta)-x_B)(\phi^P_\pi-\phi^T_\pi)\phi_v-\eta z
(\phi^P_\pi+\phi^T_\pi)\phi_v \bigg]\non
&&\times a_7(t_{1c}) E_{1cd}(t_{1c})h_{1c}(x_B,z,x_3,b_B,b_3)\non
&&+\big[\sqrt{\eta}z(\phi^P_\pi-\phi^T_\pi)((\phi_t-\phi_s)+\sqrt{\eta}\phi_v)
+(x_B-x_3(1-\eta))(\phi^P_\pi+\phi^T_\pi)\non
&&\times(\sqrt{\eta}(\phi_s+\phi_t)-\phi_v)\big]
a_7(t_{1d})E_{1cd}(t_{1d})h_{1d}(x_B,z,x_3,b_B,b_3)
\bigg\},
\eeq
and the Wilson coefficients $a_7=C_5+C_7$ and $a_9=C_3+C_9$.

For Figs.~\ref{fig-fig2}(a) and \ref{fig-fig2}(b), we have
\beq\label{DA-fig2-ab-q=u}
\mathcal{A}^{q=u}_{2(a,b)}&=&V^*_{ub}V_{ud}
F^{LL}_{B\to\pi}-V^*_{tb}V_{td}\left(F^{\prime LL}_{B\to\pi}+F^{LR}_{B\to\pi}\right),\\
\label{DA-fig2-ab-q=d}
\mathcal{A}^{q=d}_{2(a,b)}&=&-V^*_{tb}V_{td}\left(
F^{\prime \prime LL}_{B\to\pi}+F^{\prime LR}_{B\to\pi}+F^{SP}_{B\to\pi}\right).
\eeq
The amplitudes involving the $B\to\pi$ transition form factors are expressed as
\beq\label{exp-FLL-B2Pi}
F^{LL}_{B\to\pi}&=&8\pi C_F m^4_B F_{\pi}(\omega^2)\int dx_B dx_3\int b_B db_B b_3 db_3
\phi_B(x_B,b_B) (2\zeta-1)\non
&&\times\bigg\{\left[(1+x_3(1-\eta))(1-\eta)\phi^A_\pi
+r(1-2x_3)(1-\eta)\phi^P_\pi+r(1+\eta-2x_3(1-\eta))\phi^T_\pi\right]\non
&&\times a_2(t_{2a}) E_{2ab}(t_{2a})h_{2a}(x_B,x_3,b_B,b_3)\non
&&+\left[x_B(1-\eta)\eta\phi^A_\pi+2r(1-\eta(1+x_B))\phi^P_\pi
\right] a_2(t_{2b})E_{2ab}(t_{2b})h_{2b}(x_B,x_3,b_B,b_3)
\bigg\},\\
F^{LR}_{B\to\pi}&=&F^{LL}_{B\to\pi}|_{a_2\to a_6},\\
F^{\prime LL}_{B\to\pi}&=&F^{LL}_{B\to\pi}|_{a_2\to a_4}, \\
F^{\prime LR}_{B\to\pi}&=&F^{LL}_{B\to\pi}|_{a_2\to a_8},\\
F^{\prime\prime LL}_{B\to\pi}&=&F^{LL}_{B\to\pi}|_{a_2\to a_{10}}, \\
\label{exp-FSP-B2Pi}
F^{SP}_{B\to\pi}&=&16\pi C_F m^4_B\sqrt{\eta}F_{\pi}(\omega^2) \int dx_B dx_3\int b_B db_B b_3 db_3 \phi_B(x_B,b_B)\non
&&\times\bigg\{\left[(1-\eta)\phi^A_\pi+r(2+x_3(1-\eta))\phi^P_\pi
-rx_3(1-\eta)\phi^T_\pi\right]a_8^\prime(t_{2a})E_{2ab}(t_{2a})h_{2a}(x_B,x_3,b_B,b_3)\non
&&+\left[x_B(1-\eta)\phi^A_\pi+2r(1-x_B-\eta)\phi^P_\pi\right]
a_8^\prime(t_{2b})E_{2ab}(t_{2b})h_{2b}(x_B,x_3,b_B,b_3)
\bigg\},
\eeq
in which the Wilson coefficients are given by $a_2=C_1+C_2/N_c$,
$a_4=C_3+C_4/N_c+C_9+C_{10}/N_c$, $a_6=C_5+C_6/N_c+C_7+C_8/N_c$,
$a_8=C_5+C_6/N_c-C_7/2-C_8/(2N_c)$, $a_8^\prime=C_5/N_c+C_6-C_7/(2N_c)-C_8/2$,
and $a_{10}=\left[C_3+C_4-C_9/2-C_{10}/2\right](N_c+1)/N_c$.
We derive from Figs.~\ref{fig-fig2}(c) and \ref{fig-fig2}(d)
\beq\label{DA-fig2-cd-q=u}
\mathcal{A}^{q=u}_{2(c,d)}&=&V^*_{ub}V_{ud}M^{LL}_{B\to\pi}
-V^*_{tb}V_{td}\left(M^{\prime LL}_{B\to\pi}+M^{SP}_{B\to\pi}\right),\\
\label{DA-fig2-cd-q=d}
\mathcal{A}^{q=d}_{2(c,d)}&=&-V^*_{tb}V_{td}\left(M^{\prime \prime LL}_{B\to\pi}
+M^{LR}_{B\to\pi}+M^{\prime SP}_{B\to\pi}\right),
\eeq
with the amplitudes
\beq\label{exp-MLL-B2pi}
M^{LL}_{B\to\pi}&=&32\pi C_F m^4_B/\sqrt{2N_c} \int dx_B dz dx_3\int b_B db_B b db
\phi_B(x_B,b_B)\phi_v\non
&&\times\bigg\{\big[(1-x_B-z)(1-\eta^2)\phi^A_\pi+rx_3(1-\eta)(\phi^P_\pi-\phi^T_\pi)
+r(x_B+z)\eta(\phi^P_\pi+\phi^T_\pi)\non
&&-2r\eta\phi^P_\pi\big]C_2(t_{2c}) E_{2cd}(t_{2c})h_{2c}(x_B,z,x_3,b_B,b)\non
&&-\left[(z-x_B+x_3(1-\eta))(1-\eta)\phi^A_\pi+r(x_B-z)\eta(\phi^P_\pi-\phi^T_\pi)
-rx_3(1-\eta)(\phi^P_\pi+\phi^T_\pi) \right]\non
&&\times C_2(t_{2d}) E_{2cd}(t_{2d})h_{2d}(x_B,z,x_3,b_B,b)
\bigg\},\\
\label{exp-MLR-B2pi}
M^{LR}_{B\to\pi}&=&32\pi C_F m^4_B\sqrt{\eta}/\sqrt{2N_c} \int dx_B dz dx_3\int b_B db_B b db
\phi_B(x_B,b_B)\non
&&\times\bigg\{\big[(1-x_B-z)(1-\eta)(\phi_s+\phi_t)\phi^A_\pi
+r(1-x_B-z)(\phi_s+\phi_t)(\phi^P_\pi-\phi^T_\pi)\non
&&+r(x_3(1-\eta)+\eta)(\phi_s-\phi_t)(\phi^P_\pi+\phi^T_\pi)
\big]a_5^\prime(t_{2c})E_{2cd}(t_{2c})h_{2c}(x_B,z,x_3,b_B,b)\non
&&-\big[(z-x_B)(1-\eta)(\phi_s-\phi_t)\phi^A_\pi+r(z-x_B)(\phi_s-\phi_t)(\phi^P_\pi-\phi^T_\pi)\non
&&+rx_3(1-\eta)(\phi_s+\phi_t)(\phi^P_\pi+\phi^T_\pi) \big]
a_5^\prime(t_{2d}) E_{2cd}(t_{2d})h_{2d}(x_B,z,x_3,b_B,b)
\bigg\},\\
\label{exp-MSP-B2pi}
M^{SP}_{B\to\pi}&=&32\pi C_F m^4_B/\sqrt{2N_c} \int dx_B dz dx_3\int b_B db_B b db
\phi_B(x_B,b_B)\phi_v\non
&&\times\bigg\{\big[(1+\eta-x_B-z+x_3(1-\eta))(1-\eta)\phi^A_\pi
+r\eta(x_B+z)(\phi^P_\pi-\phi^T_\pi)\non
&&-rx_3(1-\eta)(\phi^P_\pi+\phi^T_\pi)
-2r\eta\phi^P_\pi\big]a_6^\prime(t_{2c}) E_{2cd}(t_{2c})h_{2c}(x_B,z,x_3,b_B,b)\non
&&-\left[(z-x_B)(1-\eta^2)\phi^A_\pi-rx_3(1-\eta)(\phi^P_\pi-\phi^T_\pi)
+r(x_B-z)\eta(\phi^P_\pi+\phi^T_\pi)\right]\non
&&\times a_6^\prime(t_{2d}) E_{2cd}(t_{2d})h_{2d}(x_B,z,x_3,b_B,b)
\bigg\},\\
M^{\prime LL}_{B\to\pi}&=&M^{LL}_{B\to\pi}|_{C_2\to a_4^\prime},\\
M^{\prime\prime LL}_{B\to\pi}&=&M^{LL}_{B\to\pi}|_{C_2\to a_{10}^\prime},\\
M^{\prime SP}_{B\to\pi}&=&M^{SP}_{B\to\pi}|_{a_6^\prime\to a_6^{\prime\prime}},
\eeq
where the Wilson coefficients are defined as $a_4^\prime=C_4+C_{10}$, $a_5^\prime=C_5-C_7/2$,
$a_6^\prime=C_6+C_8$, $a_6^{\prime\prime}=C_6-C_8/2$,
and $a_{10}^\prime=C_3+C_4-C_9/2-C_{10}/2$.

The factorizable annihilation diagrams in Figs.~\ref{fig-fig3}(a) and ~\ref{fig-fig3}(b) lead to
\beq\label{DA-fig3-ab}
\mathcal{A}_{3(a,b)}&=&V^*_{ub}V_{ud} F^{LL}_{a\pi}
-V^*_{tb}V_{td} \left(F^{\prime LL}_{a\pi}+F^{SP}_{a\pi}\right),
\eeq
with the three-pion production amplitudes
\beq\label{exp-FLL-annpi}
F^{LL}_{a\pi}&=&8\pi C_F m^4_B f_B \int dz dx_3\int b db b_3 db_3\non
&&\times\bigg\{\left[(x_3(1-\eta)-1)(1-\eta)\phi^A_\pi\phi_v
+2r\sqrt\eta(x_3(1-\eta)(\phi^P_\pi-\phi^T_\pi)-2\phi^P_\pi)\phi_s\right]\non
&&\times a_1(t_{3a})E_{3ab}(t_{3a})h_{3a}(z,x_3,b,b_3)\non
&&+\left[z(1-\eta)\phi^A_\pi\phi_v+2r\sqrt\eta\phi^P_\pi((1-\eta)(\phi_s-\phi_t)
+z(\phi_s+\phi_t))\right]\non
&&\times a_1(t_{3b})E_{3ab}(t_{3b})h_{3b}(z,x_3,b,b_3)
\bigg\},\\
F^{\prime LL}_{a\pi}&=&F^{LL}_{a\pi}|_{a_1\to a_3},\\
\label{exp-FSP-annpi}
F^{SP}_{a\pi}&=&16\pi C_F m^4_B f_B  \int dz dx_3\int b db b_3 db_3\non
&&\times\bigg\{\left[2\sqrt\eta(1-\eta)\phi^A_\pi\phi_s
+r(1-x_3)(\phi^P_\pi+\phi^T_\pi)\phi_v+r\eta((1+x_3)\phi^P_\pi-(1-x_3)\phi^T_\pi)\phi_v\right]\non
&&\times a_5(t_{3a})E_{3ab}(t_{3a})h_{3a}(z,x_3,b,b_3)\non
&&+\left[2r(1-\eta)\phi^P_\pi\phi_v+z\sqrt\eta((1-\eta)\phi^A_\pi(\phi_s-\phi_t)
+2r\sqrt\eta\phi^P_\pi\phi_v)\right]\non
&&\times a_5(t_{3b})E_{3ab}(t_{3b})h_{3b}(z,x_3,b,b_3)
\bigg\}.
\eeq
The nonfactorizable annihilation diagrams in Figs.~\ref{fig-fig3}(c) and \ref{fig-fig3}(d)
give
\beq\label{DA-fig3-cd}
\mathcal{A}_{3(c,d)}=V^*_{ub}V_{ud}M^{LL}_{a\pi}
-V^*_{tb}V_{td}\left(M^{\prime LL}_{a\pi}+M^{LR}_{a\pi} \right),
\eeq
with the amplitudes
\beq\label{exp-MLL-annpi}
M^{LL}_{a\pi}&=&32\pi C_F m^4_B/\sqrt{2N_c} \int dx_B dz dx_3\int b_B db_B b_3 db_3
\phi_B(x_B,b_B)\non
&&\times\bigg\{\big[(1-\eta)(\eta-(1+\eta)(x_B+z))\phi^A_\pi\phi_v
+r\sqrt\eta(x_3(1-\eta)+\eta)(\phi^P_\pi+\phi^T_\pi)(\phi_s-\phi_t)\non
&&-r\sqrt\eta(1-x_B-z)(\phi^P_\pi-\phi^T_\pi)(\phi_s+\phi_t)
+4r\sqrt\eta\phi^P_\pi\phi_s\big]C_1(t_{3c})E_{3cd}(t_{3c})h_{3c}(x_B,z,x_3,b_B,b_3)\non
&&+\big[(1-\eta)(1-x_3(1-\eta)-\eta(1+x_B-z))\phi^A_\pi\phi_v
-r\sqrt\eta(x_B-z)(\phi^P_\pi+\phi^T_\pi)(\phi_s-\phi_t)\non
&&+r\sqrt\eta(1-\eta)(1-x_3)(\phi^P_\pi-\phi^T_\pi)(\phi_s+\phi_t) \big]
C_1(t_{3d})E_{3cd}(t_{3d})h_{3d}(x_B,z,x_3,b_B,b_3)
\bigg\},\\
M^{\prime LL}_{a\pi}&=&M^{LL}_{a\pi}|_{C_1\to a_9},\\
\label{exp-MLR-annpi}
M^{LR}_{a\pi}&=&32\pi C_F m^4_B/\sqrt{2N_c} \int dx_B dz dx_3\int b_B db_B b_3 db_3
\phi_B(x_B,b_B)\non
&&\times\bigg\{\big[\sqrt\eta(1-\eta)(2-x_B-z)\phi^A_\pi(\phi_s+\phi_t)
-r(1+x_3)(\phi^P_\pi-\phi^T_\pi)\phi_v\non
&&-r\eta[(1-x_B-z)(\phi^P_\pi+\phi^T_\pi)-x_3(\phi^P_\pi-\phi^T_\pi)
+2\phi^P_\pi]\phi_v\big]a_7(t_{3c})E_{3cd}(t_{3c})h_{3c}(x_B,z,x_3,b_B,b_3)\non
&&-\big[r(1-\eta)(1-x_3)(\phi^P_\pi-\phi^T_\pi)\phi_v
-\sqrt\eta(x_B-z)[r\sqrt\eta(\phi^P_\pi+\phi^T_\pi)\phi_v\non
&&-(1-\eta)\phi^A_\pi(\phi_s+\phi_t)] \big]
a_7(t_{3d})E_{3cd}(t_{3d})h_{3d}(x_B,z,x_3,b_B,b_3)
\bigg\}.
\eeq

Similarly, we derive from Figs.~\ref{fig-figs}(a) and \ref{fig-figs}(b)
\beq\label{DA-fig4-ab}
\mathcal{A}_{4(a,b)}&=&V^*_{ub}V_{ud} F^{LL}_{a\pi\pi}
-V^*_{tb}V_{td}\left( F^{\prime LL}_{a\pi\pi}+F^{SP}_{a\pi\pi}
\right),
\eeq
with the three-pion production amplitudes
\beq\label{exp-FLL-apipi}
F^{LL}_{a\pi\pi}&=&8\pi C_F m^4_B f_B \int dz dx_3\int b db b_3 db_3\non
&&\times\bigg\{\left[2r\sqrt{\eta}\phi^P_\pi((2-z)\phi_s+z\phi_t)
-(1-\eta)(1-z)\phi^A_\pi\phi_v\right]a_1(t_{4a}) E_{4ab}(t_{4a})h_{4a}(z,x_3,b,b_3)\non
&&+\big[2r\sqrt{\eta}[(1-x_3)(1-z)\phi^T_\pi-(1+x_3+(1-x_3)\eta)\phi^P_\pi]\phi_s\non
&&+(x_3(1-\eta)+\eta)(1-\eta)\phi^A_\pi\phi_v \big]
a_1(t_{4b})E_{4ab}(t_{4b})h_{4b}(z,x_3,b,b_3)
\bigg\},\\
F^{\prime LL}_{a\pi\pi}&=&F^{LL}_{a\pi\pi}|_{a_1\to a_3}\\
\label{exp-FSP-apipi}
F^{SP}_{a\pi\pi}&=&16\pi C_F m^4_B f_B \int dz dx_3\int b db b_3 db_3\non
&&\times\bigg\{\left[\sqrt{\eta}(1-\eta)(1-z)\phi^A_\pi(\phi_s+\phi_t)
-2r(1+(1-z)\eta)\phi^P_\pi\phi_v \right]\non
&&\times a_5(t_{4a})E_{4ab}(t_{4a})h_{4a}(z,x_3,b,b_3)\non
&&+\left[2\sqrt{\eta}(1-\eta)\phi^A_\pi\phi_s
-r(2\eta+x_3(1-\eta))\phi^P_\pi\phi_v
+rx_3(1-\eta)\phi^T_\pi\phi_v\right]\non
&&\times a_5(t_{4b})E_{4ab}(t_{4b})h_{4b}(z,x_3,b,b_3)
\bigg\},
\eeq
and from Figs.~\ref{fig-figs}(c) and \ref{fig-figs}(d)
\beq\label{DA-fig4-cd}
\mathcal{A}_{4(c,d)}=V^*_{ub}V_{ud}M^{LL}_{a\pi\pi}
-V^*_{tb}V_{td}\left( M^{\prime LL}_{a\pi\pi}+M^{LR}_{a\pi\pi} \right),
\eeq
with the amplitudes
\beq\label{exp-MLL-apipi}
M^{LL}_{a\pi\pi}&=&32\pi C_F m^4_B/\sqrt{2N_c} \int dx_B dz dx_3\int b_B db_B b_3 db_3
\phi_B(x_B,b_B)\non
&&\times\bigg\{\big[(\eta-1)[x_3(1-\eta)+x_B+\eta(1-z)]\phi^A_\pi\phi_v
+r\sqrt\eta(x_3(1-\eta)+x_B+\eta)(\phi^P_\pi+\phi^T_\pi)\non
&&\times(\phi_s-\phi_t)+r\sqrt\eta(1-z)(\phi^P_\pi-\phi^T_\pi)(\phi_s+\phi_t)
+2r\sqrt\eta(\phi^P_\pi\phi_s+\phi^T_\pi\phi_t)\big]\non
&&\times C_1(t_{4c})E_{4cd}(t_{4c})h_{4c}(x_B,z,x_3,b_B,b_3)\non
&&+\big[(1-\eta^2)(1-z)\phi^A_\pi\phi_v
+r\sqrt\eta(x_B-x_3(1-\eta)-\eta)(\phi^P_\pi-\phi^T_\pi)(\phi_s+\phi_t)\non
&&-r\sqrt\eta(1-z)(\phi^P_\pi+\phi^T_\pi)(\phi_s-\phi_t) \big]
C_1(t_{4d}) E_{4cd}(t_{4d})h_{4d}(x_B,z,x_3,b_B,b_3)
\bigg\},\\
M^{\prime LL}_{a\pi\pi}&=&M^{LL}_{a\pi\pi}|_{C_1\to a_9},\\
\label{exp-MLR-apipi}
M^{LR}_{a\pi\pi}&=&-32\pi C_F m^4_B/\sqrt{2N_c} \int dx_B dz dx_3\int b_B db_B b_3 db_3
\phi_B(x_B,b_B)\non
&&\times\bigg\{\big[\sqrt\eta(1-\eta)(1+z)\phi^A_\pi(\phi_s-\phi_t)
+r(2-x_B-x_3(1-\eta))(\phi^P_\pi+\phi^T_\pi)\phi_v\non
&&+r\eta(z\phi^P_\pi-(2+z)\phi^T_\pi)\phi_v\big]a_7(t_{4c})E_{4cd}(t_{4c})h_{4c}(x_B,z,x_3,b_B,b_3)\non
&&+\big[\sqrt\eta(1-\eta)(1-z)\phi^A_\pi(\phi_s-\phi_t)
+r(x_3(1-\eta)-x_B)(\phi^P_\pi+\phi^T_\pi)\phi_v\non
&&+r\eta((2-z)\phi^P_\pi+z\phi^T_\pi)\phi_v \big]
a_7(t_{4d})E_{4cd}(t_{4d})h_{4d}(x_B,z,x_3,b_B,b_3)
\bigg\}.
\eeq

The threshold resummation factor $S_t(x)$
follows the parametrization in \cite{prd65-014007}
\beq\label{eq-def-stx}
S_t(x)=\frac{2^{1+2c}\Gamma(3/2+c)}{\sqrt{\pi}\Gamma(1+c)}[x(1-x)]^c,
\eeq
in which the parameter is set to $c=0.3$. The hard functions are
written as
\beq
h_{1a}(x_B,z,b_B,b)&=&K_0(m_B\sqrt{x_B z} b_B)
\big[\theta(b_B-b)K_0(m_B\sqrt{z} b_B)I_0(m_B\sqrt{z} b)+ (b \leftrightarrow b_B) \big]S_t(z),
\non
h_{1b}(x_B,z,b_B,b)&=&K_0(m_B\sqrt{x_B z} b_2)S_t(x_B)\non
& &\times\left\{ \begin{array}{ll}
\frac{i\pi}{2}\left[\theta(b-b_B)H_0^{(1)}(m_B\sqrt{\eta-x_B} b)
J_0(m_B\sqrt{\eta-x_B} b_B)+ (b \leftrightarrow b_B) \right],~x_B<\eta,\\
\left[\theta(b-b_B)K_0(m_B\sqrt{x_B-\eta} b)
I_0(m_B\sqrt{x_B-\eta} b_B)+(b\leftrightarrow b_B)\right],\quad\quad~~x_B\geq\eta,\\
\end{array} \right.
\non
h_{1c}(x_B,z,x_3,b_B,b_3)&=&\big[\theta(b_B-b_3)K_0(m_B\sqrt{x_Bz}b_B)
I_0(m_B\sqrt{x_Bz} b_3)+(b_B\leftrightarrow b_3)  \big]\non
&&\times\left\{ \begin{array}{ll}
\frac{i\pi}{2}H_0^{(1)}(m_B\sqrt{z[(1-\eta)(1-x_3)-x_B]} b_3),~\quad\quad (1-\eta)(1-x_3)> x_B,\\
K_0(m_B\sqrt{z[ x_B-(1-\eta)(1-x_3)]} b_3),~~~\quad\quad\quad(1-\eta)(1-x_3)\leq x_B,\end{array} \right.
\non
h_{1d}(x_B,z,x_3,b_B,b_3)&=&\big[\theta(b_B-b_3)K_0(m_B\sqrt{x_Bz} b_B)
I_0(m_B\sqrt{x_Bz} b_3)+(b_B\leftrightarrow b_3)  \big]\non
&&\times\left\{ \begin{array}{ll}
\frac{i\pi}{2}H_0^{(1)}(m_B\sqrt{z[x_3(1-\eta)-x_B]} b_3),~\quad\quad x_3(1-\eta)>x_B,\\
K_0(m_B\sqrt{z[x_B-x_3(1-\eta)]} b_3),~~~\quad\quad\quad x_3(1-\eta)\leq x_B,\end{array} \right.
\non
h_{2a}(x_B,x_3,b_B,b_3)&=& K_0(m_B\sqrt{x_B x_3(1-\eta)} b_B)
\big[\theta(b_B-b_3)K_0(m_B\sqrt{x_3(1-\eta)} b_B)\non
& &\times I_0(m_B\sqrt{x_3(1-\eta)} b_3)+ (b_3 \leftrightarrow b_B) \big] S_t(x_3),
\non
h_{2b}(x_B,x_3,b_B,b_3)&=&h_{2a}(x_3,x_B,b_3,b_B),
\non
h_{2c}(x_B,z,x_3,b_B,b)&=&\big[\theta(b_B-b)K_0(m_B\sqrt{x_B x_3(1-\eta)} b_B)
I_0(m_B\sqrt{x_B x_3(1-\eta)} b)\non
& &+(b_B\leftrightarrow b)  \big]\left\{ \begin{array}{ll}
\frac{i\pi}{2}H_0^{(1)}(m_B\sqrt{(1-x_B-z)[x_3(1-\eta)+\eta]} b),~\quad\quad x_B+z<1,\\
K_0(m_B\sqrt{(x_B+z- 1)[x_3(1-\eta)+\eta]} b),~~~\quad\quad\quad x_B+z\geq1,\end{array} \right.
\non
h_{2d}(x_B,z,x_3,b_B,b)&=&\big[\theta(b_B-b)K_0(m_B\sqrt{x_B x_3(1-\eta)} b_B)
I_0(m_B\sqrt{x_B x_3(1-\eta)} b)\non
& &+(b_B\leftrightarrow b)  \big]\left\{ \begin{array}{ll}
\frac{i\pi}{2}H_0^{(1)}(m_B\sqrt{x_3(z-x_B)(1-\eta)} b),~\quad\quad x_B<z,\\
K_0(m_B\sqrt{x_3(x_B-z)(1-\eta)} b),~~~\quad\quad\quad x_B\geq z,\end{array} \right.\nonumber
\non
h_{3a}(z,x_3,b,b_3)&=&\left(\frac{i\pi}{2}\right)^2H_0^{(1)}(m_B\sqrt{(1-x_3)z(1-\eta)}b)
S_t(x_3)\non
&&\times\big[\theta(b-b_3)H_0^{(1)}(m_B\sqrt{1-x_3(1-\eta)} b)
J_0(m_B\sqrt{1-x_3(1-\eta)} b_3)+ (b \leftrightarrow b_3) \big],
\non
h_{3b}(z,x_3,b,b_3)&=&\left(\frac{i\pi}{2}\right)^2H_0^{(1)}(m_B\sqrt{(1-x_3)z(1-\eta)} b_3)
S_t(z)\non
&&\times\big[\theta(b-b_3)H_0^{(1)}(m_B\sqrt{z(1-\eta)} b)
J_0(m_B\sqrt{z(1-\eta)} b_3)+ (b \leftrightarrow b_3) \big],
\non
h_{3c}(x_B,z,x_3,b_B,b_3)&=&\frac{i\pi}{2}K_0(m_B\sqrt{1-x_3(1-x_B-z)(1-\eta)+(x_B+z-1)\eta} b_B)\non
&&\times\big[\theta(b_B-b_3)H_0^{(1)}(m_B\sqrt{(1-x_3)z(1-\eta)} b_B)
J_0(m_B\sqrt{(1-x_3)z(1-\eta)} b_3)\non
&&+ (b_B \leftrightarrow b_3) \big],
\non
h_{3d}(x_B,z,x_3,b_B,b_3)&=&\frac{i\pi}{2}
\big[\theta(b_B-b_3)H_0^{(1)}(m_B\sqrt{(1-x_3)z(1-\eta)} b_B)J_0(m_B\sqrt{(1-x_3)z(1-\eta)} b_3)+ (b_B \leftrightarrow b_3) \big]\non
&&\times\left\{ \begin{array}{ll}
\frac{i\pi}{2}H_0^{(1)}(m_B\sqrt{(1-x_3)(z-x_B)(1-\eta)} b_B),~\quad\quad x_B<z,\\
K_0(m_B\sqrt{(1-x_3)(x_B-z)(1-\eta)} b_B),~~~\quad\quad\quad x_B\geq z,\end{array} \right.
\non
h_{4a}(z,x_3,b,b_3)&=&\left(\frac{i\pi}{2}\right)^2H_0^{(1)}(m_B\sqrt{(1-z)
(\eta+x_3(1-\eta))} b_3) S_t(z)\non
&&\times\big[\theta(b-b_3)H_0^{(1)}(m_B\sqrt{1-z} b)
J_0(m_B\sqrt{1-z} b_3)+ (b \leftrightarrow b_3) \big],
\non
h_{4b}(z,x_3,b,b_3)&=&\left(\frac{i\pi}{2}\right)^2H_0^{(1)}(m_B\sqrt{(1-z)
(\eta+x_3(1-\eta))} b)S_t(x_3)\non
&&\times\big[\theta(b-b_3)H_0^{(1)}(m_B\sqrt{\eta+x_3(1-\eta)}b)
J_0(m_B\sqrt{\eta+x_3(1-\eta)} b_3)+ (b \leftrightarrow b_3) \big],
\non
h_{4c}(x_B,z,x_3,b_B,b_3)&=&\frac{i\pi}{2}K_0(m_B\sqrt{1-z((1-x_3)(1-\eta)-x_B)} b_B)\non
&&\times\big[\theta(b_B-b_3)H_0^{(1)}(m_B\sqrt{(1-z)(\eta+x_3(1-\eta))} b_B)
J_0(m_B\sqrt{(1-z)(\eta+x_3(1-\eta))} b_3)\non
&&+ (b_B \leftrightarrow b_3) \big],
\non
h_{4d}(x_B,z,x_3,b,b_3)&=&\frac{i\pi}{2}
\big[\theta(b_B-b_3)H_0^{(1)}(m_B\sqrt{(1-z)(\eta+x_3(1-\eta))} b_B)\non
&&\times J_0(m_B\sqrt{(1-z)(\eta+x_3(1-\eta))} b_3)+ (b_B \leftrightarrow b_3) \big]\non
&&\times\left\{ \begin{array}{ll}
\frac{i\pi}{2}H_0^{(1)}(m_B\sqrt{(1-z)(\eta+x_3(1-\eta)-x_B)} b_B),~\quad x_B<\eta+x_3(1-\eta),\\
K_0(m_B\sqrt{(1-z)(x_B-\eta-x_3(1-\eta))} b_B),~~~\quad\quad x_B\geq \eta+x_3(1-\eta),\end{array} \right.
\eeq
with the Hankel function $H_0^{(1)}(x)=J_0(x)+iY_0(x)$.

The evolution factors in the above factorization formulas are given by
\beq
E_{1ab}(t)&=&\alpha_s(t)  \exp[-S_B(t)-S_{Ms}(t)],\non
E_{1cd}(t)&=&\alpha_s(t)  \exp[-S_B(t)-S_{Ms}(t)-S_{\pi}]|_{b=b_B},\non
E_{2ab}(t)&=&\alpha_s(t)  \exp[-S_B(t)-S_{\pi}(t)],\non
E_{2cd}(t)&=&\alpha_s(t)  \exp[-S_B(t)-S_{Ms}(t)-S_{\pi}]|_{b_3=b_B},\non
E_{3ab}(t)&=&\alpha_s(t)  \exp[-S_{Ms}-S_{\pi}(t)],\non
E_{3cd}(t)&=&\alpha_s(t)  \exp[-S_B(t)-S_{Ms}(t)-S_{\pi}]|_{b_3=b},\non
E_{4ab}(t)&=&E_{3ab}(t),\non
E_{4cd}(t)&=&E_{3cd}(t),
\eeq
in which the Sudakov exponents are defined as
\beq
S_B&=&s\left(x_B\frac{m_B}{\sqrt2},b_B\right)+\frac53\int^t_{1/b_B}\frac{d\bar\mu}{\bar\mu}
\gamma_q(\alpha_s(\bar\mu)),\non
S_{Ms}&=&s\left(z\frac{m_B}{\sqrt2},b\right)+s\left((1-z)\frac{m_B}{\sqrt2},b\right)+
2\int^t_{1/b}\frac{d\bar\mu}{\bar\mu}
\gamma_q(\alpha_s(\bar\mu)),\non
S_{\pi}&=&s\left(x_3\frac{m_B}{\sqrt2},b_3\right)+s\left((1-x_3)\frac{m_B}{\sqrt2},b_3\right)+
2\int^t_{1/b_3}\frac{d\bar\mu}{\bar\mu}
\gamma_q(\alpha_s(\bar\mu)),
\eeq
with the quark anomalous dimension $\gamma_q=-\alpha_s/\pi$. The explicit expressions
of the functions $s(Q,b)$ can be found, for example, in Appendix A of Ref.~\cite{prd76-074018}.
The involved hard scales are chosen in the PQCD approach as
\beq
t_{1a}&=&\max\left\{m_B\sqrt{z}, 1/b_B, 1/b\right\},\non
  t_{1b}&=&\max\left\{m_B\sqrt{|x_B-\eta|}, 1/b_B, 1/b\right\},\non
t_{1c}&=&\max\left\{m_B\sqrt{x_B z},m_B\sqrt{z|(1-\eta)(1-x_3)-x_B|},1/b_B, 1/b_3,\right\},\non
t_{1d}&=&\max\left\{m_B\sqrt{x_B z},m_B\sqrt{z|x_B-x_3(1-\eta)|},1/b_B, 1/b_3 \right\},\non
t_{2a}&=&\max\left\{m_B\sqrt{x_3(1-\eta)}, 1/b_B, 1/b_3\right\},\non
  t_{2b}&=&\max\left\{m_B\sqrt{x_B(1-\eta)}, 1/b_B, 1/b_3\right\},\non
t_{2c}&=&\max\left\{m_B\sqrt{x_Bx_3(1-\eta)},m_B\sqrt{|1-x_B-z|[x_3(1-\eta)+\eta]},1/b_B, 1/b,\right\},\non
t_{2d}&=&\max\left\{m_B\sqrt{x_Bx_3(1-\eta)},m_B\sqrt{|x_B-z|x_3(1-\eta)},1/b_B, 1/b \right\},\non
t_{3a}&=&\max\left\{m_B\sqrt{1-x_3(1-\eta)}, 1/b, 1/b_3\right\},\non
  t_{3b}&=&\max\left\{m_B\sqrt{z(1-\eta)}, 1/b, 1/b_3\right\},\non
t_{3c}&=&\max\bigg\{m_B\sqrt{(1-x_3)z(1-\eta)},m_B\sqrt{1-x_3(1-x_B-z)(1-\eta)+(x_B+z-1)\eta},\non
 &&\quad\quad\quad 1/b_B, 1/b_3,\bigg\},\non
t_{3d}&=&\max\left\{m_B\sqrt{(1-x_3)z(1-\eta)},m_B\sqrt{|x_B-z|(1-x_3)(1-\eta)},1/b_B, 1/b_3 \right\},\non
t_{4a}&=&\max\left\{m_B\sqrt{1-z}, 1/b, 1/b_3\right\},\non
  t_{4b}&=&\max\left\{m_B\sqrt{\eta+x_3(1-\eta)}, 1/b, 1/b_3\right\},\non
t_{4c}&=&\max\bigg\{m_B\sqrt{(1-z)(\eta+x_3(1-\eta))},m_B\sqrt{1-z((1-x_3)(1-\eta)-x_B)},\non
 &&\quad\quad\quad 1/b_B, 1/b_3,\bigg\},\non
t_{4d}&=&\max\bigg\{m_B\sqrt{(1-z)(\eta+x_3(1-\eta))},m_B\sqrt{(1-z)|x_B-\eta-x_3(1-\eta)|},\non
 &&\quad\quad\quad 1/b_B, 1/b_3 \bigg\}.
\eeq


\end{document}